# Why the relational data model can be considered as a formal basis for group operations in object-oriented systems.

Evgeniy Grigoriev (c) 2007
Grigoriev.E@gmail.com

arXiv.org ID: **0708.0361**

Relational data model defines a specification of a type "relation". However, its simplicity does not mean that the system implementing this model must operate with structures having the same simplicity. We consider two principles allowing create a system which combines object-oriented paradigm (OOP) and relational data model (RDM) in one framework. The first principle -- "complex data in encapsulated domains" -- is well known from *The Third Manifesto* by Date and Darwen. The second principle --"data complexity in names"-- is the basis for a system where data are described as complex objects and uniquely represented as a set of relations. Names of these relations and names of their attributes are combinations of names entered in specifications of the complex objects. Below, we consider the main properties of such a system.

> *"The wise man will not go over the hill -- he will pass around."*
> *Russian proverb.*

## Introduction

This paper has appeared as a reaction (but not a responce) to the paper *"Integration of programming languages with data bases: what is the problem?"* [1]. After reading it, the same perception arises that is induced by the current state of the discussed subject in total. All ideas and remarks by themselves seem true; but, as a whole, we have an image of either a labyrinth or a heap -- in general, something where you should not enter.

We will not do this -- because the problem of impedance considered in the aforementioned paper as the fundamental one casts some doubt upon its universality mainly. Impedance appears as a side effect in a quite concrete configuration which (in the most general form) includes an OO-programming system and a relational data bases management system (in quite different its implementations). One should understand that this very wide-used configuration is nevertheless a result of chaotic evolution (dependent on thousands of random factors) of interaction of the systems working with fast memory (RAM) and with slow (external) one [2]. Designers and developers of OO-languages and RDMBS during their creation met a lot of problems more important than the forthcoming problem of matched interaction [15, 16].

In this connection, it seems interesting to attract attention to possible logical alternatives of combining OOP and RDM into framework of unique system (especially as such approaches to their creation exist already). The necessity of implementation of such combined systems comes, first, form main features of OO-programming systems [3] since this is a guarantee of rich expressive capacities of the system and, second, from formality of relational data model [4, 5] (in what follows, RDM) since it is a basis of data management systems, which provide the mathematically founded group access to the data, and (in this sense) has at present no alternative.

This paper shows ability to create a system that unites these different and orthogonal conceptions (there is no need to change or/and mixture these conceptions themselves). It means an OO-system, which implements relational data model. Such a system is referred below as R⋈O-system, where the '⋈' symbol denotes the orthogonality of two conceptions.

## Relational data model. What is the obstacle we rest on?

Quite broad and indefinite representation of the term "Data Model" is met very often though it was very clearly formally defined by E. Codd [6, 17] (who is known first of all as the author of RDM). A data model is defined as a collection of types (where a type is a set of values), a collection of possible operations over instances of these types, and a collection of integrity constraints applicable to them.

In accordance with this formal definition of the notion "Data Model," the relational data model
(1) defines a type "relation",
(2) introduces operations of relational algebra (applying which to relation, we obtain relation ),
(3) defines keys, which allow us to restrict the integrity of data represented in the form of a single relation (relation keys) or several relations (external keys)

In papers devoted to problems of designing next generation DBMS, the relational data model (RDM) is quite often criticized. The usual accusations are "old", "poor system of types", "two-dimensional data representation", "atomicity of stored values", "impossibility to describe the enterprise adequately", "insufficient semantics" - the list is far from the complete one [7, 10, 11, 12, 13, 14]. When you try to delve into these papers or discuss with their authors, two ordinary fallacies appear (usually, combined).

(1) They try to use RDM as a infological tool. In other words, RDM is considered as a tool for description of the enterprise. In such cases, one usually asserts that "the enterprise must be described as a collection of relations" and reasons why such a description of the enterprise seems to be clumsy and more elegant methods are necessary.

Actually, the requirement that data should be described as a set of relations is only a condition allowing one to apply the RDM operations and constrains (on which the group data processing is based) to these data. As an analogy, I can formulate next assertion - if values are numbers, they can be applied with the arithmetical operation "addition". The latter raises no doubts; however, this does not imply that mathematics requires that information about the enterprise should be presented in the form of numbers. Similarly, the necessity of data representation in the form of relations has no links with enterprise description.

Another fact demonstrating mistakes of type (1) is the attempts to estimate "semantics" of RDM (they mean the possibility to convey or preserve a data meanings) and compare its "semantics" with other approaches. In this case, we often meet the RDM (as a formalism) to be confused with the relational scheme of data (a concrete collection of concrete relations created in terms of this formalism).

Of course, there exist neither "poor" nor "rich" "semantics" in RDM, which operates with abstract concepts. To understand this fact, we compare the phrases "there exists a relation" and "there exists a relation with the scheme (a, b, c)". In the first case, we deal with a relation in the most general sense, i.e., with a relation which satisfied a definition (given in RDM) of just this abstract relation. In the second case we deal with a relation whose header contains some concrete values (also referred as "attribute names") "a", "b", and "c." As for a relation with the scheme (Article#, Quantity, Price) its semantics looks very evidently, but, really, this is not semantics of RDM itself but only of scalar string values contained into header of relation.

(2) Relational data model is accused of all faults of systems implementing this model in a way (sometimes, to quite concrete systems). The list of these faults is exceedingly huge. One may meet sentences stating that the operations of the relational data model are "slow", "its language is inconvenient", "its system of types is poor", its domains are atomic (this is treated as a drawback), it cannot operate with complex data, etc. I do not argue, the "Relational" Data Base Management Systems available at the present time cannot be referred to as ideal; however, all these drawbacks are in neither way consequences of the relational model.

For understanding the border between the model and the system, the analogy with other areas of mathematics realized in informational systems may be useful, for instance, the analogy with the simplest arithmetic. Arithmetic operates with integer numbers by means of a set of predefined operations; in other words, a set of types (consisting of a single type "number") and a set of operations are defined in arithmetic. Recall that a "data model" is defined as a set of types, a set of operations, and a set of constraints of data integrity. Is it very similar, is not? However, it is improbable that somebody will accuse arithmetic that its operations are slow or an inconvenient language should be used in arithmetical calculations. These problems are purely implementational -- the model has no concern to them. And, of course, arithmetic cannot be "old" or "new".

In the context of this paper, the assertion that RDM cannot manipulate with complex data is of special interest.

**Data complexity in encapsulated domains. The Third Manifesto.**

The assertion that RDM has a poor system of types is an example of mistake of the second kind. It is often accompanied with lamentations about domain atomicity. Very probably these opinions are induced by the fact that the concept of domain ("atomic domain") is considered as an analogue of the simplest built-in types available in existing systems, like INTEGER or STRING.

Of course, we should remember *"The Third Manifesto"* by C.Date and H.Darwen [8] here, which proposes a possible way to creation of the next generation DBMS. A key point is that the relational model has nothing that requires restricting the domains only by simple forms. The only requirement is that values of a domain may be manipulated only by certain operations defined for this domain. Its internal representation should be encapsulated. *"The Third Manifesto"* also asserts that, in the totality of features, a class (OO-term) does not differ from a type (common term) and from domain (RDM-term). So "class" is equal to "domain" and this is the key of the union of OO- and R- in the framework of a single system.

*"The Third Manifesto"* is well known and its base ideas can be hardly disputed. However, along with domains, formal Relational Data Model has another "place" whose capacity to convey "complex data" (or, more rigorously, "data complexity"), which is possible in OO-system, is not less than that of domains. This is names.

**Data complexity in names.**

To understand what we speak about, we once again refer to arithmetic. Consider the addition operation written in arithmetic as "x+y". Since the matter is about formal constructions, "x" and "y" denote here abstract numerical values. In information system, realizing (in a way) arithmetic calculations, expression "x+y" denotes the same addition of two values (of course, numerical), but names "x" and "y" are the ones of <u>variables</u> where these values are stored in. In this simple example, these variables is probably explicitly created and named in the same context where the addition operation on their values is run. Doubtless, such a system may be considered as a good and correct realization of arithmetic (of course, if it has no trivial arithmetical errors).

But modern programming languages allow us to use much more complex and interesting expressions. For example, using expression "CurrentOrder.items.Count() + History.OrdersDetails.TotalOfCounts", we also sum two numeric values; however, the names or combinations of names used for these numbers are much more complicated. The variables containing these values are parts of much more complex structures. They do not exist outside of these structures (i.e., in contrast to variable "x," they are not created explicitly and individually). Moreover, it may occur that these variables do not exist evidently or the value is calculated.

Would somebody assert that this system realizes the arithmetic worse or less correctly than a system which may execute only "x+y," i.e., operate only with values of *explicitly created* numeric variable? Of course, no! It is important that, figuratively speaking, 2+2 produces 4. And it is absolutely unimportant where the system takes these values from, which structures hide them, how these structures are defined, how the system provides variables containing these values with names, what names and combinations of names are used -- this is internal matters of the system. From the standpoint of arithmetic, which is a formal (one may say *meaningless* -- in the sense "having no concern to the enterprise") creation, there is *no* difference between the simple *meaningless* name "x" and the expression "CurrentOrder.items.Count()" which is full of semantic for us and is defined with a complex structure present in the system.

Going back to the false idea "RDM requires that the enterprise should be described as a set of relations," let us try to imagine the very similar requirement: "arithmetic requires that the enterprise should be described as a set of numbers." Goes wrong? Let us also add the following statement "Starting from arithmetic, all data in the system should be described *only* as values of numerical variables" and "Arithmetic assumes that some numerical variables are stored, while others are calculable." Is it absurd? Yes, of course! Let us, however, for an instant imagine that people treat these absurd requirements seriously, considering them as almost axioms. Could they create OO-languages in the same form as they exist nowadays? No, they would have only "x" in this case but "CurrentOrder.items.Count()" is not possible to achieve. Unfortunately, as for the relational model, the matter is just in this way.

Recall once again that the data model consists of the following:
- a collection of types, where a type is a set of values;
- a collection of admissible operations over instances of these types, and
- a collection of integrity constraints applicable to these operations.

None words about the enterprise are present here -- the data model is formal and *meaningless*. There are also no words here about possible realizations, about variables, and a language.

So, RDM is a formal construction doubtless. It is also doubtless that we must <u>represent</u> the data (available for this user) as a set of values of the relations to use the capacity of this formal construction. However, this requirement does not imply that data must be <u>explicitly described</u> as a set of values of relations which are stored in <u>explicitly created</u> variable relations (in many realizations, they are referred to as tables) in a system realizing an RDM.

**An example of system realizing the principle "Data complexity in names": the OverRelational Manifesto.**

An example of system which (1) allows the user to <u>describe</u> data in a rather arbitrary way and (2) guarantees that in the same name space these data will be <u>represented</u> for the user as a set of relations is the R×O-system described in [9].

The R×O-system is an OO-system. <u>The main requirement</u> is that the specification of objects of the system is a set of components of the following types:
1) scalar;
2) tuple-structured types (defined on the set of scalar types);
3) relation types (also defined on the set of scalar types).

These types are named as valuable (in essence, they are analogues of "types admitting the relational assignment," defined in *"The Third Manifesto"*). The specification of an object component may contain parameters of any valuable type to specify methods.

The reference type (for which only operations of comparison, assignment, and dereferencing are defined) is classified in [9] as a scalar type, what makes possible to describe rather complex data structures. As an example, define the *specification* of a type describing some shipments.

```
CREATE CLASS SHIPMENT
{
  No INTEGER; // scalar component of type (defined on a domain) INTEGER
  WareFrom WAREHOUSE // scalar component of reference type WAREHOUSE
  Items SET OF //component of relation type
  {
    Article STRING;
    Pieces INTEGER;
   }..
  DoShip(ToShipDate DATETIME) BOOL; //scalar component taking
                                    // a scalar parameter (method specification)
}..
```

As we can see, the "shipment" is described with a value having complex structure (0NF). It is a part of a more complex structure, because it contains a reference to an object of type WAREHOUSE, the access to which can be described by the path expression "someSHIPMENT.WareFrom" (here, someSHIPMENT is a reference to an object of object type SHIPMENT). It is possible that other objects in the system contain references to the "shipment" objects. The Items component, defined as set of records with detailed information on shipped goods, is really a relation (for brevity, here and below, expressions defining the keys will be omitted).

Using RDM properties, it is possible to show that, if a system satisfies the main requirement, then data described as a set of objects having complex structure may be *represented* with the system and in the system as a set of relations. A simple naming rule allows use names and/or combinations of names introduced in description of this complex structure as names of these relations and their attribute:

*If a path expression "C.\*.\*.s", where "C" is a name determining the existence of a set of objects (it may be a class name defined in the global context or a name of a reference defined in a local context), "s" is the name of a scalar component defined in a structure present in this path expression, and "\*.\*" are the other names, which form the path expression, is correct in the system, then the system guarantees that there exists a relation with name "C.\*" in which there exists a scalar attribute named "\*.s" containing the value of component "s" and an attribute named "C" containing a reference to an object in which this component "s" is present.*

In simpler words the naming rule means that we can divide any path expression into two arbitrary parts. First part is a name of the relation; second one is a name of its attribute. Because of hierarchical structure a lot of path expressions having the same beginning can be defined; so, a lot of attributes can exist in such relation. Also this relation contains attribute referencing on the objects described by values of others attributes.

For example, suppose that the following specification exists in the system:

```
CREATE CLASS Y
{
  a scalartype;      // Here, scalartype is a scalar type (domain).
  b SET OF           // Component b is of type relation
  {                  // with attributes...
    c scalartype;    // ...c (defined on domain scalartype),
    d scalartype;    // ...d (defined on domain scalartype).
   }..
}..
```

Thus, the following structure is defined for objects of type Y (scalar elements of the structure are underlined)

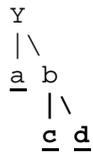

and the path expressions `"Y.a,"` `"Y.b.c,"` and `"Y.b.d"` are correct.

By virtue of the aforementioned naming rule, the R×O-system guarantees that next relations, containing data of all objects of type Y, exist and are globally accessible:

```
"Y" ("Y", "a", "b.c", "b.d")
"Y.b" ("Y", "c", "d")
```

Here the relation name and, in parentheses, its attribute names are given in quotes to show only that they are (from standpoint of formal RMD) just *meaningless* names, even if such a name looks like a part of definition of a complex structure existing in the system and is full of "semantic" for user.

Note also that if name `"refY"` of a reference to an object (or a set of objects) of type Y is defined in some context, then the system guarantees that the following relations exist in this context:

```
"refY" ("Y", "a", " b.c", " b.d")
"refY.b" ("Y", "c", "d")
```

This rule is valid for structures of any nesting depth. To illustrate this, suppose that the specification of type X exists in the system

```
CREATE CLASS X
{
  e Y;              // e is a component of the reference type Y.
  f SET OF          // Component b is of type relation
  {                 // with attribute
    g Y;            // g (defined on the reference domain Y)
  }..
}..
```

Thus, for objects of type X (with regard for the nested structure of Y), the following structure is defined:

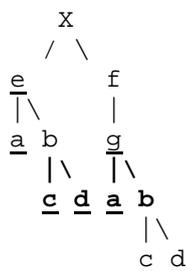

By virtue of this description, the system guarantees for the user that, along with described relations, the system has the relations

```
"X" ("X", "e", "e.a", "e.b.c", "e.b.d", "f.g", "f.g.a", "f.g.b.c", "f.g.b.d")
"X.e" ("X", "a", "b.c", "b.d")
"X.e.b" ("X", "c", "d")
"X.f" ("X", "g", "g.a", "g.b.c", "g.b.d")
"X.f.g" ("X", "a", " b.c", "b.d")
"X.f.g.b" ("X", "c", "d")
```

Thus, all data about all existing in the system objects accessible for the user (the user has access to data described in the specification of type) may be represented in the form of relations, with which the user may manipulate using the usual for relational systems operations of group access to data. A one of consequences of this fact is that it is not necessary to perform any additional actions to make group operating possible -- for instance, to create additional elements (such as ODMG's extends) which is necessary for uniting objects into groups.

Note once again that these relations are defined on the basis of *specification* of object types. Concerning type *realization*, R×O-system preserves all possibilities (intrinsic for conventional OO-programming systems) of creating and using user-defined types and opens these possibilities from a new side.

To demonstrate this fact, we come back to the SHIPMENT type
```
CREATE CLASS SHIPMENT
{
  No INTEGER;
  WareFrom WAREHOUSE;
  Items SET OF //-- and this is a set of invoice lines
  {
    Article STRING;
    Pieces INTEGER;
  }..
}..
```
This expression describes the specification of a type, i.e., an interface that may be used by the user for operating with objects of this type. In accordance with this specification, the system guarantees that, for instance, a relation "SHIPMENT.Items" ("SHIPMENT", "Article", "Pieces") exists, which may be used to create a query allowing to find article-detailed data about the total quantity of all shipped goods (SQL-like syntax is used here only for obviousness)
```
CREATE TotalsOfShippedGoods AS
SELECT si.Article,SUM(si.Pieces)
FROM Shipment.Items si
GROUP BY si.Article;
```
This query returns information from objects of type SHIPMENT existing in the system. However, these objects may be created only when a realization of t his type is defined. As a simplest example, consider the component Items realized as stored
```
ALTER CLASS SHIPMENT
REALIZE Items ...
AS STORED;
```
This realization means that component Items of objects of type SHIPMENT is represented for the user just in the same form as it is stored in the system (or, conversely, is stored in the system just in the same form s is represented). Other realizations are also possible. For example, any component may be realized as an relational operation returning a value
```
ALTER CLASS ...
REALIZE Items ...
AS
SELECT ... ;
```
…or as an algorithmic sequence of operations, which may perform more complex calculations and/or change the state of the objects.
```
ALTER CLASS Shipment …
REALIZE  DoShip(ToShipDate DATETIME) BOOL
AS
BEGIN
…
END;
```

Any realization may be changed during the type inheritance. For example, define a new type Sale describing the sale event. Of course, the sold goods must be in a way shipped from the warehouse and, therefore, we will define type Sale as a subtype of type SHIPMENT:
```
CREATE CLASS Sale EXTEND Shipment
{
  ...
  SaleItems SET OF
  {
     Article STRING;
     Pieces INTEGER;
     Price FLOAT;
  }
}
```

We must store the information about each sold commodity in detail; therefore, component `SaleItems` is realized as stored in this type.
```
ALTER CLASS Sale
REALIZE SaleItems AS STORED;
```
Inherited component `Items` defined and realized as stored in the parent class (this component represent data about shipped quantities) is re-realized as calculated now. Really, the shipped quantities can be exactly calculated from the sold quantities and there is no need to store it:
```
ALTER CLASS Sale
REALIZE Items ...
AS
SELECT Article, Sum(Pieces)
FROM SaleItems
GROUP BY Article;
```

Here, creating the subclass, we do not need to change the code which uses the specification of the parent class (it is typical in OO-systems). In particular, it is not necessary to change the above-created query `TotalShippedGoods` defined over type `Shipment`. This is due to the fact that, creating this query, we used relation `Shipment.Items`, which was later re-realized in subtype `SALE`. So, this relation contains data of a polymorphic component, which is realized in some objects as stored and in other objects as calculated (thus, this relation may be named as polymorphic too). The analogy with invoking the polymorphic methods of objects in modern OO-languages exists - the difference is that, invoking a polymorphic method, the system implicitly *chooses one* of possible realizations, while relations in R×O-systems are results of implicit *union* (relational operation UNION) of *all* possible realizations of the component of an object type, including all of its descendent types.

It is possible to show that R×O-system may be created on the basis of existing RDBMSes (an important requirement is that the language of this RDBMS contain procedure extensions) by means of a translator program, which translates the R×O-system commands into commands of the used RDBMS. Using RDM properties, the translation proposition can be proved, which may be treated as a formal justification for the possibility of group processing of information represented as a set of complex objects and group controlling of the states of these objects. By virtue of this proposition, any sequence of operations in realization of object component may be translated into a sequence of operations of the used RDBMS such that, executing last sequence, the system changes its state as if the initial sequence has been executed for each object of the defined set of objects. According to this proposition it is not necessary to use iterators evidently or implicitly (even for method execution on groups of objects).

The R×O-system can be controlled by nonprocedural language, which consists of commands for type definition, for object creation and deletion, for manipulating on the objects and for obtaining information about their states. By virtue of the aforesaid, the R×O-system may be treated as an environment of an adequate, active, long-living model of the enterprise, which is controlled by the user (a client application) by set of command and provides the user with information about its state. On the illustration below the R×O-system acts as a server part in client-server configuration.

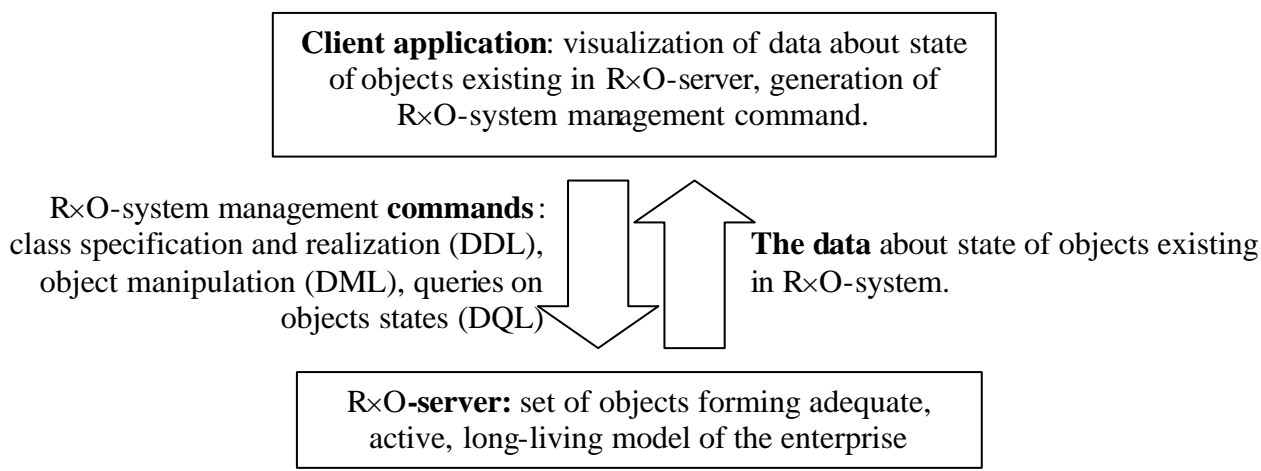

As we can see, there is no object exchange between client and server parts here; client sends only commands to manipulate objects existing in R×O-server and queries to get information about their states.

**Conclusions**

The problem of integration of databases with programming languages may hardly be solved by only technological methods. Some principal stereotypes (unfortunately, widespread) should be overcome. In particular, the relational model of data should be considered only as a formal construction.

I would like to use this possibility to thank Sergey D. Kuznetsov: this study would hardly appear without his papers and translations. I am also grateful to users of the SQL.RU site for variety of their comments which made it possible to formulate many ideas presented here.

**List of references**

Evgeniy Grigoriev (Grigoriev.E@gmail.ru), May 2007.

At present a formal grammar of an experimental management language for the described R×O -system has created, the realization of a translator for this language is in process. The author will be grateful for any propositions on cooperation in further realization of this project.